\documentclass[aps,prd,onecolumn,showpacs,showkeys,amsmath,amssymb]{revtex4}
\usepackage{graphicx}
\usepackage{dcolumn}
\usepackage{bm}
\begin{document}
\parskip 6 true pt
%
%=====================================================================================
%=====================================================================================
\title{Isospin symmetry breaking of K and K* mesons}
%=====================================================================================
%=====================================================================================
%
\author{Hua-Xing Chen}
\email{hxchen@rcnp.osaka-u.ac.jp} \affiliation{Research Center for
Nuclear Physics, Osaka University, Ibaraki 567--0047, Japan and
Department of Physics, Peking University, Beijing 100871, China}
\author{Atsushi Hosaka}
\email{hosaka@rcnp.osaka-u.ac.jp} \affiliation{Research Center for
Nuclear Physics, Osaka University, Ibaraki 567--0047, Japan}
\author{Shi-Lin Zhu}
\email{zhusl@th.phy.pku.edu.cn} \affiliation{Department of
Physics, Peking University, Beijing 100871, China}

\begin{abstract}
We use the method of QCD sum rules to investigate the isospin
symmetry breaking of K and K* mesons. The electromagnetic effect,
difference between up and down current-quark masses and difference
between up and down quark condensates are important. We perform sum
rule analyses of their masses and decay constant differences, which
are consistent with experimental values. Our results yield $\Delta
f_K = f_{K^0} - f_{K^\pm} = 1.5$ MeV.
\end{abstract}
\maketitle
\pagenumbering{arabic}

%
%=====================================================================================
%=====================================================================================
\section{Introduction}\label{sec_intro}
%=====================================================================================
%=====================================================================================
%
QCD has an approximate flavor symmetry which is determined by the
pattern of the quark masses. Isospin symmetry in particular holds to
a high accuracy. This is because the scale is set by $(m_u -
m_d)/\Lambda_\chi$, where $m$'s are current quark masses, while
$\Lambda_\chi$ is the chiral symmetry breaking scale around 1 GeV.
Because of this small hadronic isospin violations, the
electromagnetic effect becomes important in order to understand the
isospin symmetry
breaking~\cite{Das:1967it,Fujikawa:1974wa,Scorzato:2004da}. There
are many papers suggesting that the electromagnetic effect is
dominant in the mass splitting of
pions~\cite{Bardeen:1988zw,Duncan:1996xy}.

Therefore, to study the isospin symmetry breaking, it is necessary
to consider both the hadronic isospin violations and the
electromagnetic effect. In this paper, we study the isospin symmetry
breaking of $K$ and $K^*$ ($J^P=0^+$) mesons in the QCD sum rule.
This work is an extension of the previous one for the $\pi$ and
$\rho$ mesons~\cite{Zhu:1997mp}.

One can calculate the hadronic effect due to the different $up$ and
$down$ current-quark masses and condensates. While for the
electromagnetic effect, we follow the procedure in
Ref.~\cite{Kisslinger:1994me}. They constructed a gauge invariant
electromagnetic two-point function for the heavy-light quark
systems.

This paper is organized as follows. In section 2, we derive the
QCD sum rules for the $K$ and $K^*$ mesons. In section 3, we
discuss our numerical results of their masses, decay constants and
differences. We find that they are consistent with the
experimental values. Section 4 is a summary.

%
%=====================================================================================
%=====================================================================================
\section{QCD Sum Rules for $K$ and $K^*$ Mesons}\label{sec_sumrule}
%=====================================================================================
%=====================================================================================
%

For the past decades QCD sum rule has proven to be a powerful and
successful non-perturbative
method~\cite{Shifman:1978bx,Reinders:1984sr}. In sum rule analyses,
we consider two-point correlation functions:
%
%%%%%%%%%%%%%%%%%%%%%%%%%%%%%%%%%%%%%%%%%%%%%%%%%%%%%%%%%%%%%%%%%%%%%%%%%%%%%%
\begin{eqnarray}
\Pi_{\mu\nu}(q^2)\,&\equiv&\,i\int d^4x e^{iqx}
\langle0|T\eta_\mu(x){\eta_\nu^\dagger}(0)|0\rangle
\\ \nonumber &=& \Pi(q^2)(q_\mu q_\nu - q^2 g_{\mu\nu}) +
\Pi_1(q^2) q_\mu q_\nu \, ,
\label{eq_pidefine}
\end{eqnarray}
%%%%%%%%%%%%%%%%%%%%%%%%%%%%%%%%%%%%%%%%%%%%%%%%%%%%%%%%%%%%%%%%%%%%%%%%%%%%%%
%
where for $K$ meson
\begin{equation}
\eta_\mu^{(K)} = \bar q_1 \gamma_\mu \gamma_5 q_2 \, ,
\label{eq_kmeson}
\end{equation}
and for the vector $K^*$ meson
\begin{equation}
\eta_\mu^{(K^*)} = \bar q_1 \gamma_\mu q_2 \, .
\label{eq_kstarmeson}
\end{equation}
Here these currents may couple to particles $K$ and $K^*$ through
\begin{eqnarray}\nonumber
\langle 0 | \eta_\mu^{(K)}(0) | K(p) \rangle &=& i p_\mu f_K\, ,
\\ \nonumber \langle 0 | \eta_\mu^{(K^*)}(0) | K^*(p) \rangle &=& f_{K
^*}m_{K^*}\epsilon_\mu^{K^*}\, .
\end{eqnarray}
Here $p_\mu$ is the four momentum carried by the initial meson,
$f_K$ and $f_{K^*}$ are the decay constants of $K$ and $K^*$
respectively, $m_{K^*}$ is the mass of $K^*$, and
$\epsilon_\mu^{K^*}$ is the polarization vector of $K^*$. In the
OPE, $\Pi(q^2)$ can be divided into two parts: the hadronic part and
the contributions from the electromagnetic effects. The hadronic
part for $K$ and $K^*$ have been calculated in the original work of
the QCD sum rule~\cite{Shifman:1978bx,Ball:2005vx}.

\begin{figure}[hbt]
\begin{center}
\scalebox{1}{\includegraphics{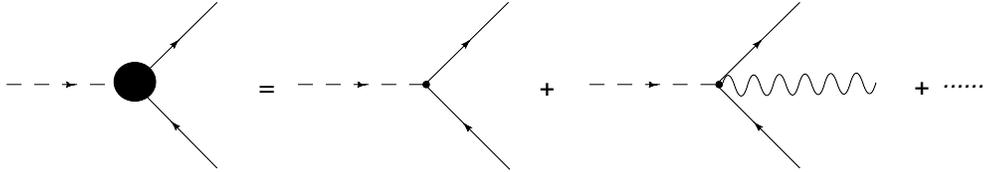}} \caption{The gauge
invariant current up to order $\alpha_e$} \label{pic_em_current}
\end{center}
\end{figure}

For the charge neutral current, like $K^0$ and $K^{*0}$, we can
change the gluons in QCD to the photons up to the order of
\mbox{$\alpha_e$ ($ \equiv e^2/4\pi $)}, and easily calculate
electromagnetic contributions. For the charged current, like $K^\pm$
and $K^{*\pm}$, the calculation of electromagnetic contributions is
slightly more complicated. If we simply change gluons to photons,
the result is not gauge invariant. To solve this problem, we follow
the procedure in Ref.~\cite{Kisslinger:1994me}. Expanding to order
$\alpha_e$, the currents become (Fig.~\ref{pic_em_current})
\begin{eqnarray}
\eta_\mu^{(K)}(q^2) &=& \bar q_1  \gamma_\mu \gamma_5  q_2 - e e_T ~
\bar q_1(q - k_1) ~~ \gamma_\mu \gamma_5 \frac{(k_1 - k_2)_\nu}{(k_1
- k_2)^2} ~~ A_\nu(k_1 - k_2) ~~ q_2(k_2) \, ,\label{eq_kmeson_em}
\\ \eta_\mu^{(K^*)}(q^2) &=& \bar q_1 \gamma_\mu q_2 - e e_T ~
\bar q_1(q - k_1) ~~ \gamma_\mu \frac{(k_1 - k_2)_\nu}{(k_1 -
k_2)^2} ~~ A_\nu(k_1 - k_2) ~~ q_2(k_2) \, ,\label{eq_kstarmeson_em}
\end{eqnarray}
where the normalized total charge of the meson is defined by $e_T
\equiv e_{q_1} - e_{q_2}$, and takes $\pm 1$ for $K^\pm$ and
$K^{*\pm}$.
%
%%%%%%%%%%%%%%%%%%%%%%%%%%%%%%%%%%%%%%%%%%%%%%%%%%%%%%%%%%%%%%%%%%%%%%%%%%%%%%
%---------figure electromagnetic_invariant_current
%%%%%%%%%%%%%%%%%%%%%%%%%%%%%%%%%%%%%%%%%%%%%%%%%%%%%%%%%%%%%%%%%%%%%%%%%%%%%%
%

We have performed the OPE calculation up to dimension six, which
contains the four-quark condensates. The results are
\begin{eqnarray}
\nonumber \Pi_{K^0 (\bar K^0)} &=& - {1 \over 4\pi^2} (1+{\alpha_s
\over \pi}) \ln {-q^2 \over \mu^2} - {1\over 4\pi^2} {e_d^2 + e_s^2
\over 2}{\alpha_e \over \pi} \ln {-q^2 \over \mu^2}
\\ && + {3 \over 8 \pi^2 q^2} (m_d + m_s)^2 \ln {-q^2 \over \mu^2}
- {1\over q^4} (m_d \langle \bar s s \rangle + m_s \langle \bar d
d\rangle )
\\ \nonumber &&
+ {1\over 12q^4}\langle {\alpha_s \over \pi} G^2 \rangle - {352 \pi
\over 81 q^6}  \alpha_s \langle \bar d d \rangle \langle \bar s s
\rangle -{44 \pi \over 27 q^6} \alpha_e e_d e_s (\langle {\bar d} d
\rangle^2 +\langle {\bar s} s \rangle^2 ) \, ,
\label{pi_neutral_kaon}
\\ \nonumber \Pi_{K^{\pm}} &=& - {1 \over 4\pi^2} (1+{\alpha_s \over
\pi}) \ln {-q^2 \over \mu^2} - {1\over 4\pi^2} e_u e_s{\alpha_e
\over \pi} \ln {-q^2 \over \mu^2}
\\ && + {3 \over 8 \pi^2 q^2} (m_u + m_s)^2 \ln {-q^2 \over \mu^2}
- {1\over q^4} (m_u \langle \bar s s \rangle + m_s \langle \bar u
u\rangle )
\\ \nonumber &&
+ {1\over 12q^4}\langle {\alpha_s \over \pi} G^2 \rangle - {352 \pi
\over 81 q^6}  \alpha_s \langle \bar u u \rangle \langle \bar s s
\rangle -{44 \pi \over 27 q^6} \alpha_e e_u e_s (\langle {\bar u} u
\rangle^2 +\langle {\bar s} s \rangle^2 ) \, ,
\label{pi_charged_kaon}
\\ \nonumber \Pi_{K^{* 0} (\bar K^{* 0})}  &=& - {1 \over 4\pi^2}
(1+{\alpha_s \over \pi}) \ln {-q^2 \over \mu^2} - {1\over 4\pi^2}
{e_d^2 + e_s^2 \over 2}{\alpha_e \over \pi} \ln {-q^2 \over \mu^2}
\\ && + {3 \over 8 \pi^2 q^2} (m_d - m_s)^2 \ln
{-q^2 \over \mu^2} + {1\over q^4} (m_d \langle {\bar s}s\rangle +
m_s \langle {\bar d}d\rangle )
\\ \nonumber && + {1\over
12q^4}\langle {\alpha_s \over \pi} G^2 \rangle + {32 \pi \over 9
q^6} \alpha_s \langle {\bar d} d \rangle \langle {\bar s} s \rangle
- {32 \pi \over 81 q^6}  \alpha_s ( \langle {\bar d} d \rangle^2 +
\langle {\bar s} s \rangle^2)
\\ \nonumber &&
+ {28 \pi \over 27 q^6} \alpha_e e_d e_s (\langle {\bar d} d
\rangle^2 + \langle {\bar s} s \rangle^2)\,
,\label{pi_neutral_kstar}
\\ \nonumber \Pi_{K^{*\pm}} &=& - {1 \over 4\pi^2} (1+{\alpha_s \over \pi}) \ln
{-q^2 \over \mu^2} - {1\over 4\pi^2} e_u e_s{\alpha_e \over \pi} \ln
{-q^2 \over \mu^2}
\\ && + {3 \over 8 \pi^2 q^2} (m_u - m_s)^2 \ln
{-q^2 \over \mu^2} + {1\over q^4} (m_u \langle {\bar s}s\rangle +
m_s \langle {\bar u}u\rangle )
\\ \nonumber && + {1\over
12q^4}\langle {\alpha_s \over \pi} G^2 \rangle + {32 \pi \over 9
q^6} \alpha_s \langle {\bar u} u \rangle \langle {\bar s} s \rangle
- {32 \pi \over 81 q^6} \alpha_s ( \langle {\bar u} u \rangle^2 +
\langle {\bar s} s \rangle^2)
\\ \nonumber && + {\pi \over
q^6} \alpha_e e_T^2 \langle {\bar u} u \rangle \langle {\bar s} s
\rangle + {28 \pi \over 27 q^6} \alpha_e e_u e_s (\langle {\bar u} u
\rangle^2 + \langle {\bar s} s \rangle^2)\, .
\label{pi_charged_kstar}
\end{eqnarray}
In these equations, $u,d,s$ represent $u$, $d$ and $s$ quarks
respectively. The couplings $e_u$, $e_d$ and $e_s$ are normalized by
the unit electric charge $e$, and therefore, $e_u = 2/3$ and $e_d =
e_s = -1/3$. The quantities $\langle \bar{u}u \rangle$, $\langle
\bar{d}d \rangle$ and $\langle \bar{s}s \rangle$ are dimension $D=3$
quark condensates, and $\langle g^2 GG \rangle$ is a $D=4$ gluon
condensate. We have assumed the vacuum dominance and factorization
for the four quark condensates, for instance~\cite{Shifman:1978bx},
\begin{eqnarray}
\nonumber && \langle 0 | \bar q \gamma_\mu \gamma_5 {\lambda^a \over
2} q \bar q \gamma^\mu \gamma_5 {\lambda^a \over 2} q | 0 \rangle =
{16 \over 9} \langle 0 | \bar q q | 0 \rangle ^2 \, ,
\\ \nonumber && \langle 0 | \bar q \sigma_{\mu\nu} \gamma_5 {\lambda^a \over
2} q \bar q \sigma^{\mu\nu} \gamma_5 {\lambda^a \over 2} q | 0
\rangle = {16 \over 3} \langle 0 | \bar q q | 0 \rangle ^2 \, .
\end{eqnarray}

The difference \mbox{$\Pi_{K^0} - \Pi_{K^\pm}$} determines the
isospin symmetry breaking of $K$ meson, while the difference
\mbox{$\Pi_{K^{*0}} - \Pi_{K^{*\pm}}$} determines the isospin
symmetry breaking of $K^*$ meson. If we consider that the difference
between the $up$ and $down$ quark condensates is small and introduce
the average condensate \mbox{$\langle \bar q q \rangle = ( \langle
\bar u u \rangle + \langle \bar d d \rangle ) / 2$}, we find
\begin{eqnarray}\nonumber
\Pi_{K^0} - \Pi_{K^\pm}  &\approx& {1 \over 8 \pi^2} ( 2 e_u e_s -
e_d^2 - e_s^2 ) {\alpha_e \over \pi} \ln{ -q^2 \over \mu^2 } - {3
\over 8\pi^2
q^2}( m_u + m_d + 2 m_s ) ( m_u - m_d ) \ln{ -q^2 \over \mu^2 } \\
\nonumber && + {1 \over q^4} (m_u - m_d) \langle \bar s s \rangle
+{1 \over q^4} m_s (\langle \bar u u \rangle - \langle \bar d d
\rangle ) \\ \nonumber && + { 352 \pi \over 81 q^6 } \alpha_s
\langle \bar s s \rangle ( \langle \bar u u \rangle  - \langle \bar
d d \rangle ) + { 44 \pi \over 27 q^6 } \alpha_e ( e_u - e_d ) e_s (
\langle \bar q q \rangle^2 + \langle \bar s s \rangle^2 )\, ,
\\ \nonumber \Pi_{K^{*0}} - \Pi_{K^{*\pm}}
&\approx& {1 \over 8 \pi^2} ( 2 e_u e_s - e_d^2 - e_s^2 ) {\alpha_e
\over \pi} \ln{ -q^2 \over \mu^2 } - {3 \over 8\pi^2
q^2}( m_u + m_d - 2 m_s ) ( m_u - m_d ) \ln{ -q^2 \over \mu^2 } \\
\nonumber && - {1 \over q^4} (m_u - m_d) \langle \bar s s \rangle -
{1 \over q^4} m_s (\langle \bar u u \rangle - \langle \bar d d
\rangle ) - { 32 \pi \over 9 q^6 } \alpha_s  \langle \bar s s
\rangle
( \langle \bar u u \rangle  - \langle \bar d d \rangle ) \\
\nonumber && + { 32 \pi \over 81 q^6 } \alpha_s  ( \langle \bar u u
\rangle ^2  - \langle \bar d d \rangle ^2 ) - {\pi \over q^6}
\alpha_e e_T^2 \langle \bar u u \rangle \langle \bar s s \rangle - {
28 \pi \over 27 q^6 } \alpha_e ( e_u - e_d ) e_s ( \langle \bar q q
\rangle^2 + \langle \bar s s \rangle^2 )\, .
\end{eqnarray}

There are three non-perturbative effects
%
%%%%%%%%%%%%%%%%%%%%%%%%%%%%%%%%%%%%%%%%%%%%%%%%%%%%%%%%%%%%%%%%%%%%%%%%%%%%%%
\begin{enumerate}

\item
The difference due to the masses of $up$ and $down$ quarks.

\item
The difference between the $up$ and $down$ quark condensates.

\item
The electromagnetic part containing four-quark condensates which are
of the first order of $\alpha_e$.

\end{enumerate}
%%%%%%%%%%%%%%%%%%%%%%%%%%%%%%%%%%%%%%%%%%%%%%%%%%%%%%%%%%%%%%%%%%%%%%%%%%%%%%
%

The difference between the $up$ and $down$ quark condensates has
been evaluated previously. We define $\lambda$ to be
\begin{equation}
\lambda \equiv { \langle \bar d d \rangle \over \langle \bar u u
\rangle } - 1
\end{equation}
For instance, Gasser and Leutwyler obtained $\lambda \approx
-0.0074$~\cite{Gasser:1984gg}, while in Ref~\cite{Hatsuda:1990pj},
Hatsuda, Hogaasen and Prakash found $-0.0078 \lesssim \lambda
\lesssim -0.0067$. In the QCD sum rule, Chernyak and Zhitnitsky
obtained $\lambda \approx -0.009$~\cite{Chernyak:1983ej}. Here we
will use the value $\lambda \approx -0.0074$.

If we choose $q^2 \sim m_K^2$, the above three effects are in the
same order of magnitude. This is different from the $\pi$ and $\rho$
mesons, where only the electromagnetic part
dominates~\cite{Zhu:1997mp}.

Within the approximation of the narrow resonance with a continuum
above threshold value $s_0$, after the Borel transformation, we
obtain the final QCD sum rules
\begin{eqnarray}
\nonumber f^2_{K^0} e^{ -{m^2_{K^0} \over M_B^2} } &=& {1\over
4\pi^2} (1+{\alpha_s \over \pi} + e_d e_s{\alpha_e \over \pi} )
M_B^2 (1- e^{- {s_0\over M_B^2}}) - {3 \over 4 \pi^2} (m_d + m_s)^2
\ln {M_B} (1- e^{- {s_0\over M_B^2}})
\\ && - {1 \over M_B^2} ( m_d \langle \bar s s \rangle + m_s \langle \bar d d \rangle
) + {1 \over 12 M_B^2}\langle {\alpha_s \over \pi} G^2 \rangle
\\ \nonumber && + {176 \pi \over 81 M_B^4} \alpha_s {\langle \bar d d \rangle \langle \bar s s \rangle}
+ {22 \pi \over 27 M_B^4} \alpha_e e_d e_s (\langle \bar d d \rangle
^2 + \langle \bar s s \rangle ^2) \, , \label{br_neutral_kaon}
\\ \nonumber
f^2_{K^{\pm}} e^{ -{m^2_{K^{\pm}} \over M_B^2} } &=&  {1\over
4\pi^2} (1+{\alpha_s \over \pi} + e_u e_s{\alpha_e \over \pi} )
M_B^2 (1- e^{- {s_0\over M_B^2}}) - {3 \over 4 \pi^2} (m_u + m_s)^2
\ln {M_B} (1- e^{- {s_0\over M_B^2}})
\\ && - {1 \over M_B^2} ( m_u \langle \bar s s \rangle + m_s \langle \bar u u \rangle
) + {1 \over 12 M_B^2}\langle {\alpha_s \over \pi} G^2 \rangle
\\ \nonumber && + {176 \pi \over 81 M_B^4} \alpha_s {\langle \bar u u \rangle \langle \bar s s \rangle}
+ {22 \pi \over 27 M_B^4} \alpha_e e_u e_s (\langle \bar u u \rangle
^2 + \langle \bar s s \rangle ^2) \, , \label{br_charged_kaon}
\\ \nonumber f^2_{K^{*0}} e^{ -{m^2_{K^{*0}} \over M_B^2} } &=&  {1\over 4\pi^2}
(1+{\alpha_s \over \pi} + e_d e_s{\alpha_e \over \pi} ) M_B^2 (1-
e^{- {s_0\over M_B^2}}) - {3 \over 4 \pi^2} (m_d - m_s)^2 \ln {M_B}
(1- e^{- {s_0\over M_B^2}})
\\ && + {1 \over M_B^2} ( m_d \langle \bar s s \rangle + m_s \langle \bar d d \rangle
) + {1 \over 12 M_B^2}\langle {\alpha_s \over \pi} G^2 \rangle - {16
\pi \over 9 M_B^4} \alpha_s {\langle \bar d d \rangle \langle \bar s
s \rangle}
\\ \nonumber && +  {16 \pi \over 81 M_B^4} \alpha_s (\langle \bar d
d \rangle ^2 + \langle \bar s s \rangle ^2)  - {14 \pi \over 27
M_B^4} \alpha_e e_d e_s (\langle \bar d d \rangle ^2 + \langle \bar
s s \rangle ^2) \, , \label{br_charged_kstar}
\\ \nonumber f^2_{K^{*\pm}} e^{ -{m^2_{K^{*\pm}} \over M_B^2} } &=&  {1\over
4\pi^2} (1+{\alpha_s \over \pi} + e_u e_s{\alpha_e \over \pi} )
M_B^2 (1- e^{- {s_0\over M_B^2}}) - {3 \over 4 \pi^2} (m_u - m_s)^2
\ln {M_B} (1- e^{- {s_0\over M_B^2}})
\\ && + {1 \over M_B^2} ( m_u \langle \bar s s \rangle + m_s \langle \bar u u \rangle
) + {1 \over 12 M_B^2}\langle {\alpha_s \over \pi} G^2 \rangle - {16
\pi \over 9 M_B^4} \alpha_s {\langle \bar u u \rangle \langle \bar s
s \rangle}
\\ \nonumber && +  {16 \pi \over 81 M_B^4} \alpha_s (\langle \bar u
u \rangle ^2 + \langle \bar s s \rangle ^2)  - {14 \pi \over 27
M_B^4} \alpha_e e_u e_s (\langle \bar u u \rangle ^2 + \langle \bar
s s \rangle ^2) +  { \pi \over 2 M_B^4} \alpha_e e_T^2 (\langle \bar
u u \rangle \langle \bar s s \rangle) \, . \label{br_charged_kstar}
\end{eqnarray}

%
%=====================================================================================
%=====================================================================================
\section{Numerical Results}\label{sec_numerical}
%=====================================================================================
%=====================================================================================
%

For numerical calculations, we use the following values of
condensates~\cite{Gasser:1984gg,Eidelman:2004wy,Yang:1993bp,Gimenez:2005nt,Jamin:2002ev,Ioffe:2002be,Ovchinnikov:1988gk}:
%
%%%%%%%%%%%%%%%%%%%%%%%%%%%%%%%%%%%%%%%%%%%%%%%%%%%%%%%%%%%%%%%%%%%%%%%%%%%%%%
\begin{eqnarray}
\nonumber &&\langle \bar q q \rangle = { \langle \bar u u \rangle +
\langle \bar d d \rangle \over 2 } = -(0.240 \mbox{ GeV})^3\, ,
\\ \nonumber && \lambda = { \langle \bar d d \rangle \over \langle \bar u u
\rangle } - 1 = -0.0074
\\ && \langle\bar ss\rangle=(0.8\pm 0.1)\times\langle \bar q q \rangle^3\, ,
\\
\nonumber &&\langle {\alpha_s \over \pi} G^2 \rangle = 0.012 \mbox{
GeV}^4\, ,
\\ \nonumber && m_u = 5.3 \mbox{MeV},~m_d = 9.4 \mbox{MeV},~m_s=130 \mbox{MeV}\,
,
\\ \nonumber && \alpha_e = {1 \over 137},~\alpha_s = {0.7}\, .
\end{eqnarray}
%%%%%%%%%%%%%%%%%%%%%%%%%%%%%%%%%%%%%%%%%%%%%%%%%%%%%%%%%%%%%%%%%%%%%%%%%%%%%%
%
The $up$ and $down$ quark condensates have uncertainly in the
absolute values. However, we keep their difference $\lambda \approx
-0.0074$.

%
%=====================================================================================
%=====================================================================================
\subsection{The QCD Sum Rule for the $K$ meson}\label{subsec_kaon}
%=====================================================================================
%=====================================================================================
%

Differentiating Eqs.~(\ref{br_neutral_kaon}) and
(\ref{br_charged_kaon}) with respect to $\frac{1}{M_B^2}$ and
dividing the results by themselves, we obtain the masses for $K^0$
($\bar K^0$) and $K^\pm$ mesons. For the study of the QCD sum
rule, we have two parameters, the threshold value $s_0$ and the
Borel mass $M_B$. Herein below we study the Borel mass dependence
in the region $0.5 \lesssim M_B^2 \lesssim 3.0$ GeV$^2$ with $s_0
\sim 0.9$ GeV$^2$. Further discussions on the $s_0$ dependence
will be presented in the end of this work.

For the absolute values of the mass of the $K$ meson, the present
QCD sum rule does not work well, because $K$ is the
Nambu-Goldstone boson having a strong collective nature due to the
non-perturbative QCD dynamics. However, the isospin symmetry
breaking effects can reasonably be studied in the QCD sum rule.

\begin{figure}[hbt]
\begin{center}
\scalebox{0.8}{\includegraphics{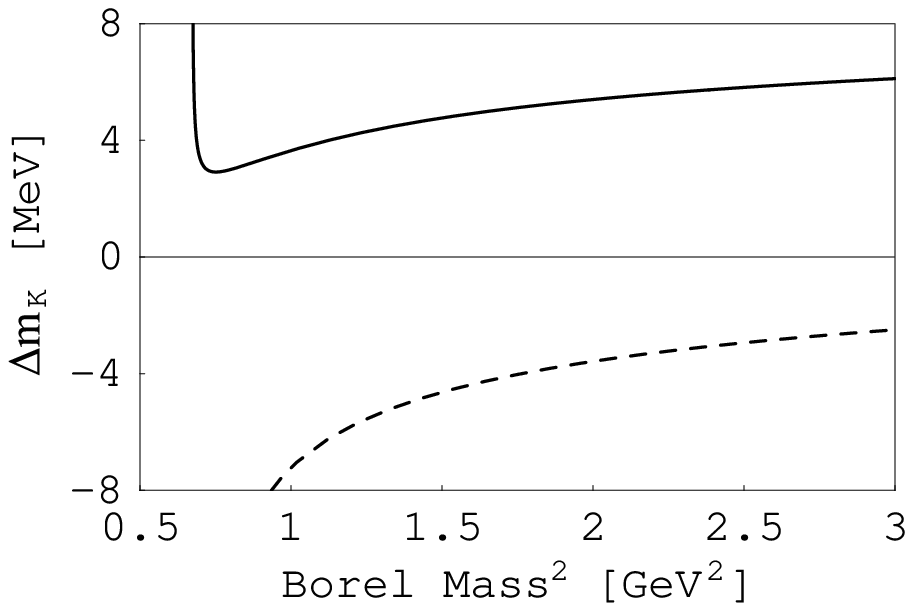}} \caption{The mass
difference of the $K$ meson, as a function of the Borel mass square
$M_B^2$. The dashed curve is obtained when \mbox{$s_0(K^0, \bar K^0)
= s_0(K^\pm) = 0.900$ GeV$^2$}. The solid curve is obtained when
\mbox{$s_0(K^0, \bar K^0) = 0.916$ GeV$^2$} and \mbox{$s_0(K^\pm) =
0.900$ GeV$^2$}.} \label{pic_kaon_mass}
\end{center}
\end{figure}

The mass difference ($\Delta m_K = m_{K^0 (\bar K^0)}-m_{K^\pm}$) is
shown in Fig.~\ref{pic_kaon_mass} as a function of the Borel mass
square $M_B^2$. The dashed curve is obtained when the threshold
value \mbox{$s_0 = 0.900$ GeV$^2$} is used both for $K^0$ ($\bar
K^0$) and $K^\pm$. The resulting mass difference turns out to be
negative which does not agree with the experiment. Also the Borel
stability is not good. We can fine tune the threshold value $s_0$
and use different values for $K^0$ ($\bar K^0$) and $K^\pm$. The
solid line is obtained when we take \mbox{$s_0(K^0, \bar K^0) =
0.916$ GeV$^2$} and \mbox{$s_0(K^\pm) = 0.900$ GeV$^2$}, with which
the sum rule value takes $\Delta m_K = 4 \pm 1$ MeV ($M_B^2 \gtrsim
1$ GeV$^2$). This is consistent with the experimental value $\Delta
m_K = 3.972 \pm 0.027$ MeV~\cite{Eidelman:2004wy}. The Borel
stability is also improved for $M_B^2 \gtrsim 1$ GeV$^2$.

%
%%%%%%%%%%%%%%%%%%%%%%%%%%%%%%%%%%%%%%%%%%%%%%%%%%%%%%%%%%%%%%%%%%%%%%%%%%%%%%
%---------figure Mass of Kaon
%%%%%%%%%%%%%%%%%%%%%%%%%%%%%%%%%%%%%%%%%%%%%%%%%%%%%%%%%%%%%%%%%%%%%%%%%%%%%%
%

\begin{figure}[hbt]
\begin{center}
\scalebox{1}{\includegraphics{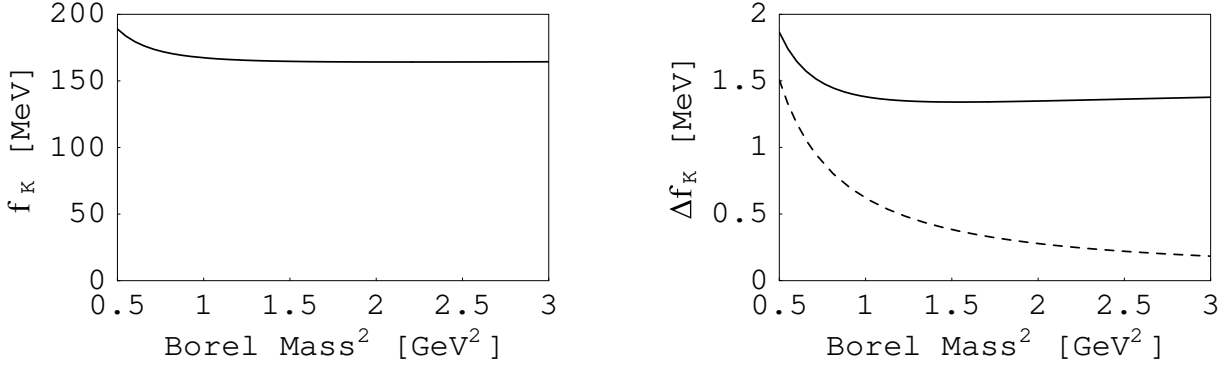}} \caption{The left
panel shows the $K^\pm$ decay constant $f_{K^\pm}$, as a function of
the Borel mass square $M_B^2$, for threshold value $s_0 = 0.900$
GeV$^2$. The right panel shows the difference of the $K$ decay
constants, as a function of the Borel mass square $M_B^2$. The
dashed curve is obtained when \mbox{$s_0(K^0, \bar K^0) = s_0(K^\pm)
= 0.900$ GeV$^2$}. The solid curve is obtained when \mbox{$s_0(K^0,
\bar K^0) = 0.916$ GeV$^2$} and \mbox{$s_0(K^\pm) = 0.900$
GeV$^2$}.} \label{pic_kaon_decay}
\end{center}
\end{figure}

Now let us study the $K$ decay constant. We need to input the mass
of the $K$ meson which we use the experimental values, $m_{K^0 (\bar
K^0)} = 497.6$ MeV and $m_{K^\pm} = 493.7$
MeV~\cite{Eidelman:2004wy}. The left panel of
Fig.~\ref{pic_kaon_decay} shows the $K^\pm$ decay constant
$f_{K^\pm}$ as a function of the Borel mass square $M_B^2$ when $s_0
= 0.900$ GeV$^2$ is used. The result for $K^0$ ($\bar K^0$) and
$K^\pm$ can not be distinguished in this figure (see the right panel
and discussion below). It is interesting that the sum rule values
take around $165$ MeV with a good Borel stability and is consistent
with the experimental value $f_{K^\pm} = 159.8 \pm 1.84$
MeV~\cite{Eidelman:2004wy}.

The difference of the $K$ decay constants $\Delta f_K = f_{K^0 (\bar
K^0)} - f_{K^\pm}$ is plotted in the right panel of
Fig.~\ref{pic_kaon_decay}, as a function of the Borel mass square
$M_B^2$. The meaning of the dashed and solid curves are the same as
for Fig.~\ref{pic_kaon_mass}. When the same threshold values are
used, $\Delta f_K$ takes values $0.2 \sim 0.5$ MeV for $1 \lesssim
M_B^2 \lesssim 3$ GeV$^2$ with some strong Borel mass dependence.
However, when using the different threshold values, we obtain
$\Delta f_K = 1.5 \pm 0.2$ MeV with a good Borel stability for $1
\lesssim M_B^2 \lesssim 3$ GeV$^2$.

%
%%%%%%%%%%%%%%%%%%%%%%%%%%%%%%%%%%%%%%%%%%%%%%%%%%%%%%%%%%%%%%%%%%%%%%%%%%%%%%
%---------figure decay constant of Kaon
%%%%%%%%%%%%%%%%%%%%%%%%%%%%%%%%%%%%%%%%%%%%%%%%%%%%%%%%%%%%%%%%%%%%%%%%%%%%%%
%

%
%=====================================================================================
%=====================================================================================
\subsection{The QCD Sum Rule for the $K^*$ Meson}\label{subsec_kstar}
%=====================================================================================
%=====================================================================================
%

\begin{figure}[hbt]
\begin{center}
\scalebox{1}{\includegraphics{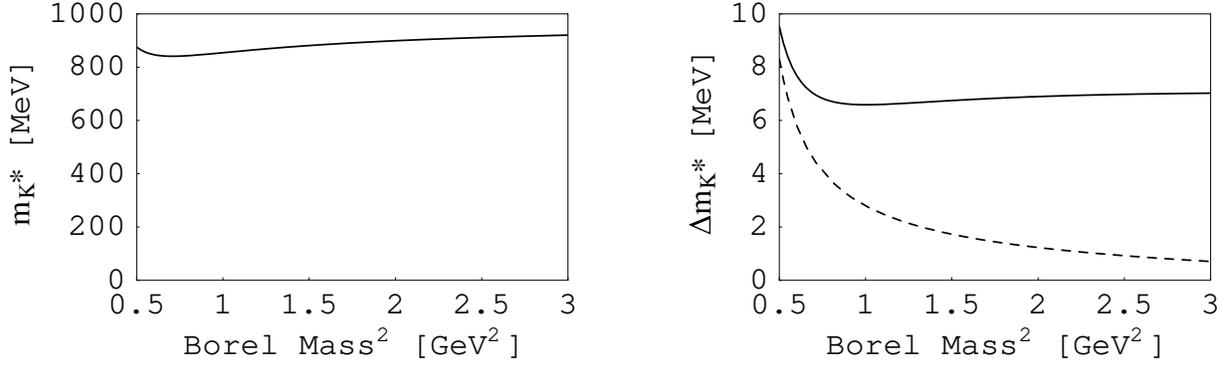}} \caption{The masses
of $K^{*0}$($\bar K^{*0}$) and $K^{*\pm}$ and their differences, as
a function of the Borel mass square $M_B^2$ for threshold values
$s_{K^{*0}, \bar K^{*0}} = 1.83$ GeV$^2$ and $s_{K^{*\pm}} = 1.80$
GeV$^2$.} \label{pic_kstar_mass}
\end{center}
\end{figure}

For the $K^*$ meson, we expect that the QCD sum rule works well just
as in the case of the $\rho$ meson. In order to check the validity
of the present sum rule, we show the mass of $K^{*\pm}$ in the left
panel of Fig.~\ref{pic_kstar_mass}, where we find a very good Borel
stability. The absolute value depends slightly on the choice of the
threshold value $s_0$, which we choose $s_0 = 1.80$ GeV$^2$ to
reproduce the experimental value $m_{K^{*\pm}} = 891.7$
MeV~\cite{Eidelman:2004wy}. The result for $K^{*0}$ ($\bar K^{*0}$)
is very similar.

The mass difference $\Delta m_{K^*} = m_{K^{*0}} - m_{K^{*\pm}}$ is
shown in the right panel of Fig.~\ref{pic_kstar_mass} as a function
of the Borel mass square $M_B^2$. The dashed curve is obtained when
the same threshold value \mbox{$s_0 = 1.80$ GeV$^2$} is used both
for $K^{*0}$ ($\bar K^{*0}$) and $K^{*\pm}$. The Borel stability is
not good. We can fine tune the threshold value $s_0$ again and use
different ones for $K^{*0}$ ($\bar K^{*0}$) and $K^{*\pm}$. The
solid line is obtained when we take \mbox{$s_0(K^{*0}, \bar K^{*0})
= 1.83$ GeV$^2$} and \mbox{$s_0(K^{*\pm}) = 1.80$ GeV$^2$}, with
which the sum rule value takes $\Delta m_{K^*} = 7 \pm 1$ MeV
($M_B^2 \gtrsim 1$ GeV$^2$). This is consistent with the
experimental value $\Delta m_{K^*} = 6.7 \pm 1.2$
MeV~\cite{Eidelman:2004wy}. The Borel stability is much improved for
$M_B^2 \gtrsim 1$ GeV$^2$.

%
%%%%%%%%%%%%%%%%%%%%%%%%%%%%%%%%%%%%%%%%%%%%%%%%%%%%%%%%%%%%%%%%%%%%%%%%%%%%%%
%---------figure Mass of Kstar
%%%%%%%%%%%%%%%%%%%%%%%%%%%%%%%%%%%%%%%%%%%%%%%%%%%%%%%%%%%%%%%%%%%%%%%%%%%%%%
%

\begin{figure}[hbt]
\begin{center}
\scalebox{1}{\includegraphics{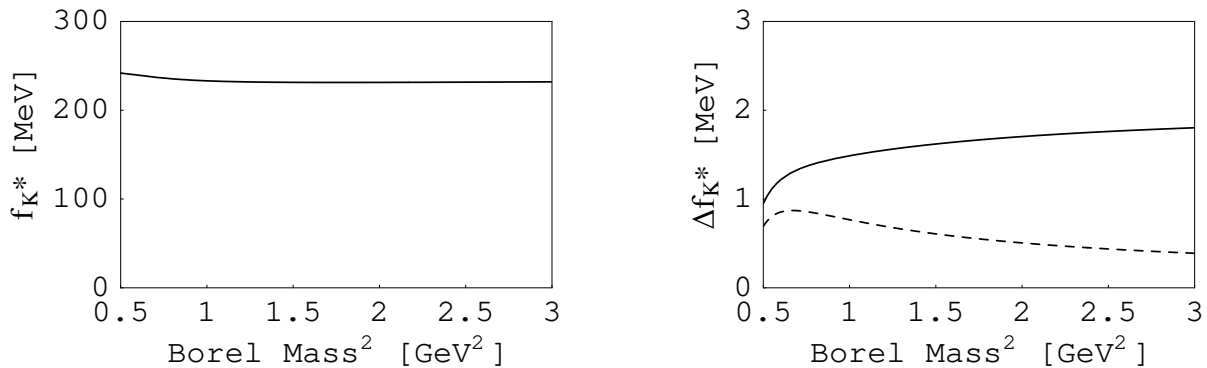}} \caption{The decay
constants of $K^{*0}$($\bar K^{*0}$) and $K^{*\pm}$ and their
differences, as a function of the Borel mass square $M_B^2$ for
threshold values $s_{K^{*0}, \bar K^{*0}} = 1.83$ GeV$^2$ and
$s_{K^{*\pm}} = 1.80$ GeV$^2$.} \label{pic_kstar_decay}
\end{center}
\end{figure}

For the study of $K^*$ decay constant, we use the experimental
values for the mass of $K^*$ meson, $m_{K^{*0}} = 896.1$ MeV and
$m_{K^{*\pm}} = 891.7$ MeV~\cite{Eidelman:2004wy}. The left panel of
Fig.~\ref{pic_kstar_decay} shows the $K^{*\pm}$ decay constant
$f_{K^{*\pm}}$ as a function of the Borel mass square $M_B^2$ when
$s_0 = 1.80$ GeV$^2$ is used. The result for $K^{*0}$ ($\bar
K^{*0}$) is very similar. The sum rule values take around $230$ MeV
with a good Borel stability. We can estimate the decay rate of
$\tau^- \rightarrow K^{*-}\nu_\tau$ as~\cite{Donoghue:1992}
\begin{eqnarray}\nonumber
\Gamma_{\tau \rightarrow K^{*-}\nu_\tau} &=& |V_{us}|^2 {G_\mu^2
\over 8 \pi} \left ( f_{K^*} \over \sqrt{2} m_{K^*}\right )^2
m^3_\tau m^2_{K^*} \left ( 1 - {m^2_{K^*} \over m^2_\tau } \right
)^2 \left ( 1 + 2 {m^2_{K^*} \over m^2_\tau } \right )
\\ \nonumber &=& 3.42 \times 10^{-14} \mbox{ GeV}\, ,
\end{eqnarray}
which is not far from the experimental values $\Gamma_{\tau^-
\rightarrow K^{*-}\nu_\tau} = 2.78 \times 10^{-14}$
GeV~\cite{Eidelman:2004wy}.

We can also calculate the decay rate of $\tau^- \rightarrow
K^{-}\nu_\tau$
\begin{eqnarray}\nonumber
\Gamma_{\tau \rightarrow K^{-}\nu_\tau} &=& |V_{us}|^2 {G_\mu^2
\over 8 \pi} \left ( f_{K} \over \sqrt{2} m_{K}\right )^2 m^3_\tau
m^2_{K} \left ( 1 - {m^2_{K} \over m^2_\tau } \right )^2
\\ \nonumber &=& 1.67 \times 10^{-14} \mbox{ GeV}\, ,
\end{eqnarray}
which is also consistent with the experimental values
$\Gamma_{\tau^- \rightarrow K^{-}\nu_\tau} = 1.48 \times 10^{-14}$
GeV~\cite{Eidelman:2004wy}.

The difference of the $K^*$ decay constants $\Delta f_{K^*} =
f_{K^{*0} (\bar K^{*0})} - f_{K^{*\pm}}$ is plotted in the right
panel of Fig.~\ref{pic_kstar_decay}, as a function of the Borel mass
square $M_B^2$. The meaning of the dashed and solid curves are the
same as for Fig.~\ref{pic_kstar_mass}.

%
%%%%%%%%%%%%%%%%%%%%%%%%%%%%%%%%%%%%%%%%%%%%%%%%%%%%%%%%%%%%%%%%%%%%%%%%%%%%%%
%---------figure decay constant of Kstar
%%%%%%%%%%%%%%%%%%%%%%%%%%%%%%%%%%%%%%%%%%%%%%%%%%%%%%%%%%%%%%%%%%%%%%%%%%%%%%
%

%
%=====================================================================================
%=====================================================================================
\subsection{The Threshold Value $s_0$ Dependence}\label{subsec_threshold}
%=====================================================================================
%=====================================================================================
%

\begin{figure}[hbt]
\begin{center}
\scalebox{1}{\includegraphics{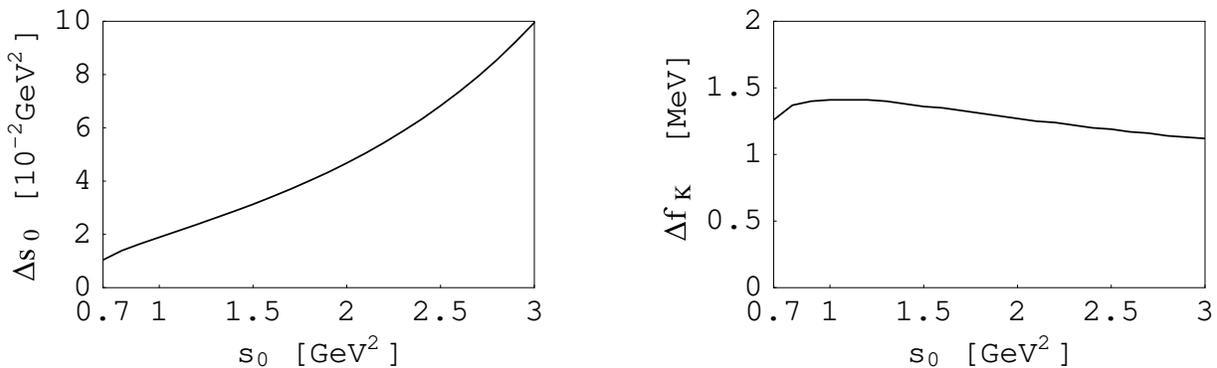}} \caption{$\Delta s_0
\equiv s_0(K^0) - s_0(K^\pm)$ and $\Delta f_K$ as functions of
$s_0$. $\Delta s_0$ is determined so as to reproduce the
experimental value of $\Delta m_K = 3.972$ MeV at $M_B^2=1$
GeV$^2$.}\label{pic_kaon}
\end{center}
\end{figure}

\begin{figure}[hbt]
\begin{center}
\scalebox{1}{\includegraphics{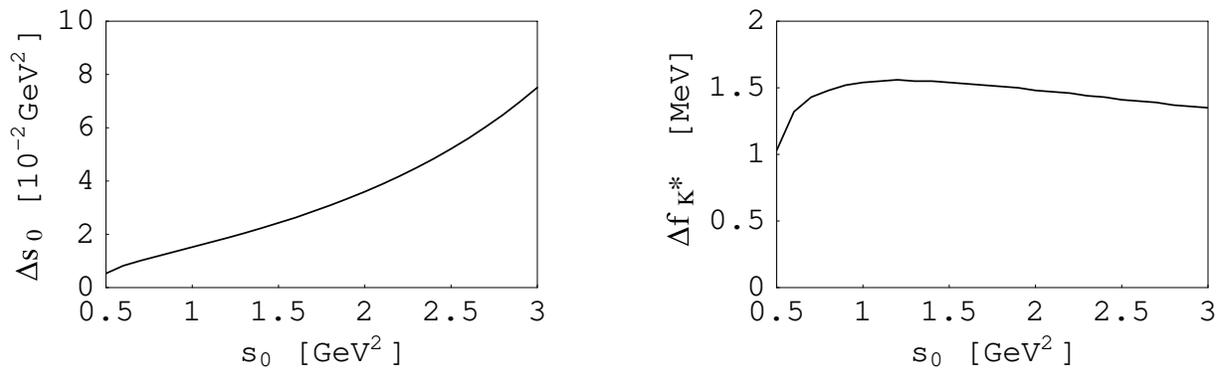}} \caption{$\Delta s_0
\equiv s_0(K^{*0}) - s_0(K^{*\pm})$ and $\Delta f_{K^*}$ as
functions of $s_0$. $\Delta s_0$ is determined so as to reproduce
the experimental value of $\Delta m_{K^*} = 6.7$ MeV at $M_B^2=1$
GeV$^2$.}\label{pic_kstar}
\end{center}
\end{figure}

Finally, let us investigate $s_0$ dependence of the present QCD sum
rule analyses. In order to see its typical behavior, we fix the
Borel mass to be $M_B^2=1$ GeV$^2$. We use the two threshold values
\begin{eqnarray}
\nonumber s_0(K^0,K^{*0}) &=& s_0 + \Delta s_0, \mbox{ ~~~~~for
$K^0$ or $K^{*0}$}\, ,
\\ \nonumber s_0(K^\pm,K^{*\pm}) &=& s_0,
\mbox{ ~~~~~~~~~~~~~~for $K^\pm$ or $K^{*\pm}$}\, .
\end{eqnarray}
As $s_0$ is varied, $\Delta s_0$ is fixed such that the experimental
mass difference $\Delta m_K$ or $\Delta m_{K^*}$ is reproduced. The
differences of the decay constants $\Delta f$ are then computed as
functions of $s_0$. The resulting $\Delta s_0$ and $\Delta f$ are
plotted in Fig.~\ref{pic_kaon} for $K$ and in Fig.~\ref{pic_kstar}
for $K^*$. It is interesting to observe that although $\Delta s_0$
are monotonically increasing functions, $\Delta f$'s are rather
stable as $s_0$ is varied. It would be an indication that the
present sum rule analyses especially for $\Delta f$ are stable.

%
%%%%%%%%%%%%%%%%%%%%%%%%%%%%%%%%%%%%%%%%%%%%%%%%%%%%%%%%%%%%%%%%%%%%%%%%%%%%%%
%---------K with n-s0
%%%%%%%%%%%%%%%%%%%%%%%%%%%%%%%%%%%%%%%%%%%%%%%%%%%%%%%%%%%%%%%%%%%%%%%%%%%%%%
%
%
%%%%%%%%%%%%%%%%%%%%%%%%%%%%%%%%%%%%%%%%%%%%%%%%%%%%%%%%%%%%%%%%%%%%%%%%%%%%%%
%---------K with n-s0
%%%%%%%%%%%%%%%%%%%%%%%%%%%%%%%%%%%%%%%%%%%%%%%%%%%%%%%%%%%%%%%%%%%%%%%%%%%%%%
%

%
%=====================================================================================
%=====================================================================================
\section{Summary}
%=====================================================================================
%=====================================================================================
%

In this paper, we have studied isospin breaking for masses and
decay constants of $K$ and $K^*$. We have adopted gauge invariant
currents coupled by a photon field. We have then estimated isospin
symmetry breaking effects through different values of the
parameters such as quark masses, condensates and threshold values.
Quark masses and condensates were fixed from other sources, while
the threshold values were fixed such that the mass differences of
charged and neutral  $K$ and $K^*$ were reproduced.  The resulting
decay constants were found to be very stable against the change in
the Borel mass and the threshold values. The resulting values for
$\Delta m$ and $\Delta f$ are consistent with experimental values.

The present analysis with good stability indicates that the QCD
sum rule can be applied to study the symmetry breaking effects in
hadron physics. In the near future, BESIII collaboration will
measure the mass splittings of $K$ and $K^\ast$ systems precisely.
Investigation of isospin symmetry breaking patterns helps to
explore the low-energy sector of the underlying QCD dynamics.

%
%=====================================================================================
%=====================================================================================
%=====================================================================================
\section*{Acknowledgments}
%=====================================================================================
%=====================================================================================
%=====================================================================================
%
H.X.C is grateful to the Monkasho fellowship for supporting his
stay at Research Center for Nuclear Physics where this work is
done. A.H. is supported in part by the Grant for Scientific
Research ((C) No.16540252) from the Ministry of Education,
Culture, Science and Technology, Japan. S.L.Z. was supported by
the National Natural Science Foundation of China under Grants
10375003 and 10421503, Ministry of Education of China, FANEDD, Key
Grant Project of Chinese Ministry of Education (NO 305001) and SRF
for ROCS, SEM.

%
%%%%%%%%%%%%%%%%%%%%%%%%%%%%%%%%%%%%%%%%%%%%%%%%%%%%%%%%%%%%%%%%%%%%%%%%%%%%%%

\end{document}